# From inference to prediction: how machine learning is reconfiguring science

Malena Méndez Isla[1], ORCID https://orcid.org/0000-0002-9748-2709

Vincent Larivière*[1,2,3,4], ORCID https://orcid.org/0000-0002-2733-0689

Diego Kozlowski[1], ORCID https://orcid.org/0000-0002-5396-3471

1 Chaire UNESCO sur la science ouverte, École de bibliothéconomie et des sciences de l'information, Université de Montréal, Montréal, Québec, Canada

2 Consortium Érudit, Montréal, Québec, Canada

3 Observatoire des sciences et des technologies, Centre interuniversitaire de recherche sur la science et la technologie, Université du Québec à Montréal, Montréal, Québec, Canada

4 Department of Science and Innovation, National Research Foundation Centre of Excellence in Scientometrics and Science, Technology and Innovation Policy, Stellenbosch University, Stellenbosch, South Africa

* Corresponding author

**Email:** malena.mendez.isla@umontreal.ca; vincent.lariviere@umontreal.ca; diego.kozlowski@umontreal.ca

## Abstract

Artificial intelligence (AI) is reshaping scientific practices, yet its epistemic implications remain underanalyzed. While recent advances in large language models are substantial, machine learning (ML) has a deeper history across disciplines. This manuscript examines 4.9 million publications and 255 ML techniques to understand how the latter are reconfiguring scientific methods and knowledge production. Through embedding-based mapping, we reconstructed the semantic space of ML research, and found a core–periphery structure where physical sciences form the methodological core and health sciences represent the primary area of adoption. Methodological profiles vary by domain: predictive techniques are concentrated in computer sciences, while inferential approaches remain distributed across applied fields. Predictive architectures, however, are displacing inference-oriented techniques in domains that have traditionally prioritized interpretability, such as health and social sciences. This displacement unfolds in two distinct waves: first (2015–2021), through deep learning architectures that reduced predictive error at the expense of epistemic opacity; and second (post-2022), through reliance on external commercial models that introduce further layers of opacity over inaccessible data and processes. This transformation expands science's analytical capacity and reshapes how knowledge is produced and evaluated.

## Introduction

Machine learning (ML) is reshaping epistemic practices. Within life sciences, for example, AlphaFold solved the 50-year-old protein structure prediction problem (Jumper et al., 2021). Before the eruption of large language models (LLMs), convolutional neural networks (CNNs) had already reached specialist-level performance in clinical image diagnosis (Esteva et al., 2017). A further illustration is that transformer-based topic models such as BERTopic (Grootendorst, 2022) have displaced latent Dirichlet allocation (Blei et al., 2003) as the standard for large-scale text analysis (Benz et al., 2025). This expansion gained international scientific recognition in 2024 when the Nobel Prizes in Physics and Chemistry were awarded to artificial intelligence researchers.

Apart from being a technical expansion of analytical capacity, the rise of AI constitutes an epistemological shift in how knowledge is generated and validated within science (Kitchin, 2014), which creates a tension between interpretability and opacity. Privileging high-performance, opaque models over less accurate but interpretable models has epistemological implications for what a discipline accepts as a valid explanation (Rudin, 2019). While the predictive accuracy of deep learning has been proven, the opacity of these architectures makes them difficult to interpret.

As Lipton (2018) points out, interpretability in ML does not have a single definition. It arises from the need to understand and validate automated decisions that impact lives in critical areas such as healthcare, justice, and ethics. Lipton (2018) distinguishes two categories of interpretability: transparency, which concerns the model's internal mechanisms (how does the model work?), and post-hoc explanations, which provide visual or textual justifications of results without requiring full mechanistic understanding. The adoption of complex architectures hinders this transparency, either due to technical opacity resulting from the mismatch between mathematical optimization and human reasoning, or commercial opacity, where secrecy surrounding code, data (Hjørland, 2026), and weights prevent their evaluation (Burrell, 2016) and reproducibility (Binz et al., 2025). While post-hoc explanations have encouraged a broad research agenda on explainable AI and  (XAI) and human-centered explainable AI (Ridley, 2025), we restrict the scope of interpretability to the category of transparency in this paper. In doing so, we focus on the potential to know and comprehend the internal mechanisms of models, a goal currently hindered by the dual nature of opacity described above. The adoption of opaque models gives rise to what Silva et al. (2024, p. 755) refer to as an "intellectual debt": the integration of systems whose logic is unknown. This reflects a structural tension between the transparency goals of open science and the opacity of AI algorithms (Borgman & Brand, 2024).

AI encompasses computational approaches that execute tasks historically performed by human cognition, no matter if it attempts to represent the biological processes that underpin them (Russell & Norvig, 2022). ML is a set of practices for achieving AI in which algorithms are trained on data to reach goals through automatic pattern recognition (Hastie et al., 2017; Jordan & Mitchell, 2015). ML is, therefore, a central component of current AI techniques and, as we show in our results, works as the overarching category that encompasses various families of algorithms that range from linear regression to deep learning (DL). DL employs artificial neural network structures with multiple layers (a deep architecture of connected artificial neurons) (LeCun et al., 2015). It is within DL that transformers (Vaswani et al., 2017) were developed, an architecture that gave rise to large language models (LLMs), which are pre-trained models on large corpora that can be repurposed for multiple tasks (Brown et al., 2020).

Scientific discourse is highly situated and rarely refers to AI in general: experts typically use precise technical terms to refer to their methods (Munk et al., 2024). This makes retrieval based on concrete techniques more faithful. Thus, this study focuses on specific ML techniques mentioned in titles or abstracts to account for the development and use of AI in contemporary science. Ding et al. (2026) analyzed the role of AI in scientific publications, differentiating between four modes: foundational (theoretical advances in AI), adaptation (modifications of existing models), tool (use of AI as an applied tool), and discussion (critical or conceptual analysis of AI). They found that tool-related papers remain the dominant category across domains. We distinguish between two levels of ML tools. At the instrumental level, LLMs are being used to write academic papers (Cabanac et al., 2021; Cabanac & Labbé, 2021; Kousha & Thelwall, 2026) or code. At the methodological level, ML composes a set of different techniques that are the pathways for answering research questions (Marradi et al., 2007). Each technique carries epistemological assumptions and has embedded validation criteria that define legitimate knowledge within a specific field. In this paper, we focus on the methodological level, and consider ML as an array of techniques, architectures and applications that comprise our taxonomy of 255 subcategories. We distinguish between techniques (e.g., logistic regression) and architectures (e.g., CNN) where analytically relevant, particularly in the context of the inference/prediction divide developed below.

Breiman (2001) identified two cultures in statistical modeling: the data modeling culture, where transparent models are a prerequisite for legitimate knowledge, and the algorithmic culture, where predictive accuracy is the primary criterion. This tension is a historical trade-off (James et al., 2023) between inference, where the goal is to understand the mechanisms behind a phenomenon, and prediction, where the aim is to estimate outcomes without the need to understand those mechanisms. While linear models facilitate interpretable inference they often lack the predictive power of nonlinear methods, such as deep learning (James et al., 2023).

As predictive techniques see exponential growth and classical statistical models decline, we ask whether these disciplinary practices are being held. To examine this methodological shift, we address four research questions:

1. How have the volume and disciplinary distribution of ML publications evolved between 1990 and 2025?
2. What is the semantic organization of ML research across scientific fields?
3. How are inference and prediction-oriented techniques distributed across that semantic space, and what does this show about the epistemic logics that differentiate knowledge validation across disciplines?
4. Has the rise of deep learning architectures displaced inference-oriented techniques in domains where transparency has historically been a condition for knowledge production? If so, to what extent?

Existing bibliometric analyses on AI can be classified according to their scope. While some papers study the application of AI within specific fields of research, others study AI as a field in itself, focused on the evolution of specific methodological families. A third body of research adopts a comprehensive perspective on the AI landscape.

At the level of AI-applications within disciplines, Shi et al. (2023) found that AI-related publications indexed by PubMed increased 36-fold from 2000 to 2022. As to natural sciences, Hao et al. (2026) discovered that AI expands personal impact, reduces the need for younger researchers, and narrows research topics toward areas already rich in data. Su et al. (2021) observed that in engineering, machine learning literature grew an average of 24.3% annually, with DL and random forests as central research topics. In management, Patra et al. (2024) identified ML, neural networks, and blockchain as the main research topics, alongside applications in sustainability and business auditing. Bircan and Salah (2022) found that AI and big data publications in social sciences are more oriented to computational studies than to social science concepts, concerns, and challenges (Bircan & Salah, 2022).

AI is often studied as a field in itself, with analyses focused on the evolution of specific methodological families, such as deep learning (Li et al., 2020), neural networks (Bianchini et al., 2022), and GenAI (Dwivedi & Elluri, 2024). Li et al. noticed DL field grew sharply after 2015, with medical imaging emerging as the dominant application domain (Li et al., 2020). Bianchini et al. (2022) found that neural network adoption grew exponentially and tended to have more citations than non-AI papers. In GenAI, Dwivedi and Elluri (2024) found that research focuses on the development of deep neural networks for image processing and computer vision. Tang et al. (2024) documented that the disciplinary and topical diversity of research teams promote innovation diffusion in the field of AI.

A third group of studies has developed a broader perspective on AI; identifying the different techniques used (OECD, 2020), or providing a meta-analysis of AI over a long period of time (Hajkowicz et al., 2023). Hajkowicz et al. (2023) tracked the diffusion of AI across 333 research fields between 1960 and 2021, and showed that its absorption reached more than 98% of fields by 2021. AI has evolved over the past decade from a specialized area of research in computer sciences into an emerging tool across all major academic disciplines (Hunt & Crawley, 2026). Ding et al. (2025) compared the diffusion of GenAI vs. other AIs (2017-2023), where GenAI is associated with larger teams and international collaboration. AI integration is positively associated with both scientific disruption and novelty, particularly within STEM disciplines (Liu et al., 2025). Duede et al. (2024) showed 13-fold growth in AI research (1985-2022) and a semantic tension where AI homogenizes production toward the style of computer sciences. Now, Zhao and Li (2026) showed that prestige positively correlates with citing highly visible AI work, while experience shows the opposite pattern both within and beyond computer science contexts. The OECD (2020) developed an operational definition of AI based on scientific publications, open-source software, and patents, and documents a sustained acceleration in AI-related scientific production since the mid-2000s, with an annual growth of 23% since 2015.

Taken together, these studies show that the adoption of AI in science is a structural change with epistemic implications. However, more research is needed to study AI as a conglomerate of multiple and diverse machine learning techniques across all fields of knowledge to analyze the methodological change in science. This manuscript aims to fill this gap.

The rest of the paper is organized as follows. In the methodology section, we first present our taxonomy and data retrieval strategy, followed by our methods, including the embedding-based semantic mapping and the inference-prediction framework. The results section addresses each of the four research questions in turn. We close with a discussion of the epistemic implications of the observed results, followed by a conclusion that summarizes our main findings.

## Methodology

### Taxonomy and data retrieval

To map the diffusion of ML across scientific fields, we developed a hierarchical taxonomy that serves as the foundation for our data retrieval strategy. Our taxonomy is based on the OECD (2020) list, which has an initial set of 186 ML keywords. We subsequently refined and extended this list into a hierarchical structure to capture both historical roots and the most recent developments in ML. Based on the definition of ML as automatic pattern recognition, we include linear and logistic regressions. These classical statistical techniques are at the foundation of many supervised models and remain essential validation standards across diverse scientific disciplines (Bzdok et al., 2018). Including these techniques allows for a better understanding of methodological change as we represent part of our results disaggregated by technique. We also added more recent techniques, such as BERT, ChatGPT, graph neural networks, transformer architectures, and vision transformers, to reflect frontier of machine learning research. Our taxonomy maps research across three main branches: supervised and unsupervised ML, deep learning, and broad family descriptors and general terminology (e.g., artificial intelligence). The full taxonomy of 255 keywords is provided in Appendix S1.

We used the OpenAlex database accessed through the InSySPo platform (Mazoni & Costas, 2024). To identify the corpus, we operationalized our taxonomy by using the 255 previously defined ML techniques as search queries. An article was included in the dataset if at least one of these techniques appeared in its title or abstract. This process yielded a corpus of 4,945,090 academic publications spanning from 1990 to 2025. The dataset has no spatial restrictions.

### Methods

Our analyses are structured along two dimensions: machine learning techniques and disciplinary boundaries. OpenAlex disciplines are divided into four levels: 4 domains, 26 fields, 244 subfields, and 4,516 research topics, of which 4,453 are represented in our corpus with at least one paper (98.6%). The research topics allow us to have a granular aggregation of papers. We restrict the analysis to the main topic of each paper.

Using the multilingual E5 (mE5) large-instruct model (Wang et al., 2024), we mapped the semantic structure of the corpus. Each publication was projected into a 1024-dimensional embedding space, where the distance between two vectors represents their thematic similarity (Wang et al., 2024). For the aggregated representation of entities (domains, fields, topics, techniques) we computed the average of the embeddings of all articles comprising that entity. To visualize this high-dimensional space, we first applied PCA to reduce the initial 1,024-dimensional embeddings to a 50-components representation, and used UMAP (McInnes et al., 2020) to project these components into a two-dimensional continuous space. In this map (Figures 2 and 3), proximity between two entities (such as two fields) indicates high semantic similarity. We used Gemini 2.5 Pro for code development and manually refined and validated all outputs.

To ensure the cross-disciplinary validity of our findings, we addressed the issue of inter-field polysemy: instances where keywords share a common token that represent different meanings across disciplines. Following Méndez Isla et al. (2025), we excluded the term neural network from our analysis of neurosciences to avoid confusion between biological brain networks and artificial architectures. Similarly, GPT was excluded from the field of Pharmacology, where the acronym traditionally refers to Glutamate Pyruvate Transaminase, a liver toxicity biomarker, rather than Generative Pre-trained Transformers (Pant, 2024).

## Analytical framework: inference vs. prediction

We operationalized the trade-off between interpretability and predictive precision, and labeled the ML techniques, according to their epistemic goal, in three categories: inference, prediction, and other. Inference measures specific variables to achieve interpretability and the main instances are linear and logistic regression. Prediction is exemplified by deep learning architectures (e.g., CNN, RNN) and complex ensembles (e.g., boosting, bagging) that reduce statistical error through non-linear opaque models. Table 1 presents this framework applied to a selection of techniques from our taxonomy. The complete list of keywords and their assignments is provided in Table S1.

**Table 1**

Classification of machine learning and deep learning techniques by type of problem they solve and epistemic goal. The selection criteria for the techniques were based on some of the techniques mentioned previously and in the results and represent a subset of the 255 techniques of our taxonomy.

| Family | Broader | Type problem | Technique | Epistemic goal |
|---|---|---|---|---|
| Machine learning | Supervised | classification | logistic regression | inference |
| | | regression | linear regression | inference |
| | | classification & regression | Boosting | prediction |
| | | | Bagging | prediction |
| | Unsupervised | dimensionality reduction | principal component analysis | inference |
| | Deep learning | classification & regression | neural network | prediction |
| | | | deep neural network | prediction |
| | | | recurrent neural networks | prediction |
| | | computer vision | convolutional neural networks | prediction |

**Results**

**Disciplinary composition and temporal growth of the ML corpus**

First, we aim to understand the prevalence of ML across domains and its evolution over time. Figure 1 shows the evolution of the corpus (Panel A), the relative weight of ML within each domain (Panel B), and its distribution across fields (Panel C). Panel A shows steady growth from 1990 until approximately 2015, followed by an exponential acceleration. Panel B shows that during the 1990s ML represented a small fraction (less than 2%) of any domain. In 2015 there was a jump, stronger in physical sciences and also notable in other domains, that corresponds with the deep learning spike that was after continued by the LLMs spike.

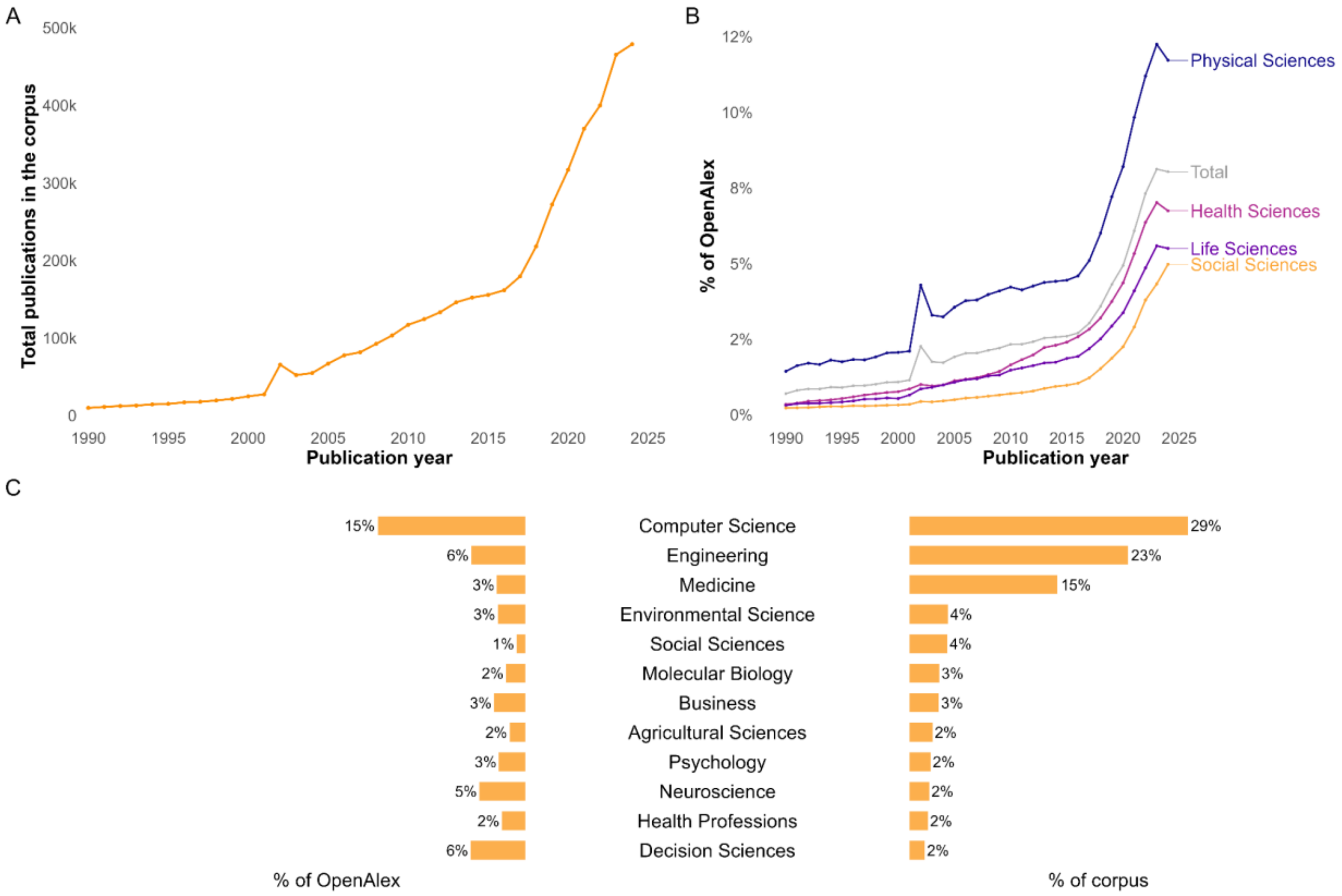

**Figure 1.** (A) Absolute temporal evolution of the corpus. (B) Relative share of ML publications in OpenAlex by domain and in total over time. (C) Relative share of ML publications in OpenAlex by field (left X axis) and percentage of publications in ML corpus per field (right X axis). Data for 2002 is influenced by Crossref reindexing effects and the inclusion of backlogged 1990s proceedings. These metadata inconsistencies in OpenAlex suggest caution when interpreting that specific year.

Panel C (left X axis) shows the absorption of ML per field in OpenAlex, that represents 15% of computer sciences, 6% of engineering, and 3% of medicine. Other high-absorption fields are decision sciences (6%) and neurosciences (5%). Panel C (right X axis) illustrates the disciplinary distribution of our corpus. As expected, the composition is heavily anchored in the fields responsible for technical development. Computer sciences (28%) and engineering (23%) concentrate over half of total production and are followed by medicine (15%), which stands as the primary field of application. The remaining 23 disciplines, including environmental sciences (4%) and social sciences (4%), form a vast and fragmented long tail, representing the progressive absorption of ML into domains where it is applied.

## ML semantic space

Figure 2 maps topics and fields according to the content of their papers’ titles and abstracts (see Methods section). The spatial proximity between topics and fields represent their semantic similarity, while the absolute orientation in the axis is arbitrary in UMAP. Topics and fields are coloured according to the domain. The domains of physical sciences (blue) and health sciences (pink) are well defined with contours reflecting regions of high concentration. Physical sciences is the densest and most expanded region, here we have the two main fields: computer sciences and engineering that are the core of ML research. Computer sciences is the largest field and its topics are expanded across the map. Other physical sciences fields have a lower density, as they are positioned outside the contour.

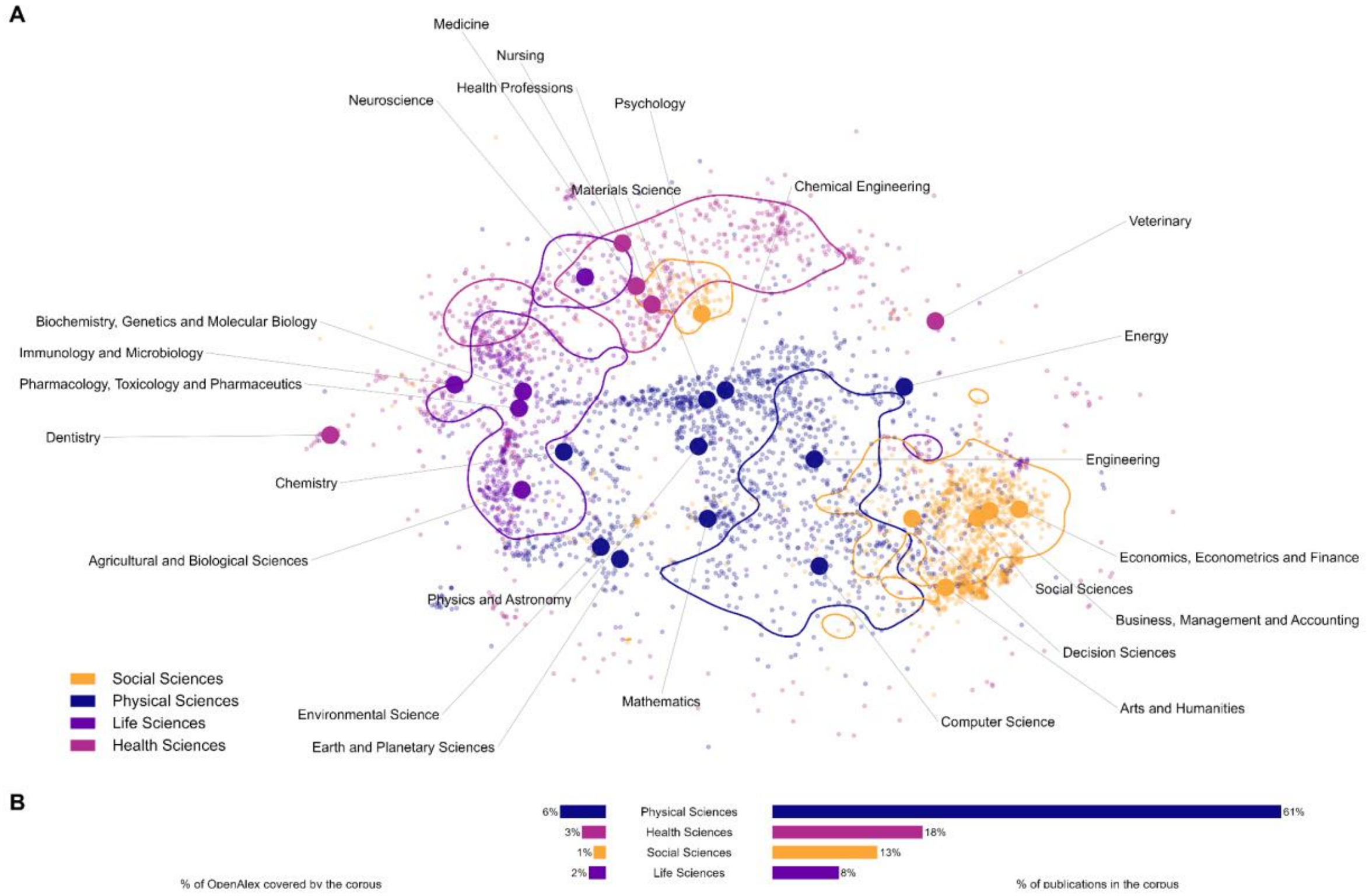

**Figure 2.** (A) ML semantic space: UMAP projection of topics and fields (1990-2025). Each smaller point represents the semantic barycenter of a topic and each larger point, the semantic barycenter of a field. This barycenter is calculated as the average of the embeddings of all articles associated with that topic and field. Contour lines represent a weighted two-dimensional kernel density estimation (KDE) of topics, computed for each domain and weighted by the number of documents associated with each topic. The contours highlight regions of higher concentration of papers within each domain. (B) For each domain (left) the percentage of OpenAlex publications covered by the corpus and (right) the domain's share within the total corpus.

Social sciences have their own cluster with the greater semantic homogeneity in the right side of the map. While psychology is labeled as part of social sciences by OpenAlex, when it comes to papers applying ML techniques, they seem to be closer to health sciences. Thus, we can observe a second cluster of social sciences topics next to health sciences.

The convergence observed between the social sciences and the physical sciences suggests that, for ML, both domains operate on abstract and formalized objects. In these fields, the focus is on the systemic and relational behavior of the object, not on its substance. In contrast, in health and life sciences, techniques deal with biological and molecular entities (such as proteins or tissues) whose material and chemical constraints impose a limit on abstraction. While in social and physical sciences, ML operates on the structure of information, in health and life sciences it operates on the complexity of living matter. Ultimately, this spatial distribution suggests that the distance in the ML semantic space is influenced by the structure of the data problem, a factor deeply intertwined with the methodological choices of each discipline.

Panel B shows that health sciences is the second largest domain, and medicine is the third largest field, with the highest absorption of ML techniques similar to engineering (as Figure 1 panel C details). Life sciences is the domain with the lowest ML absorption.

**Inference and prediction in the ML semantic space**

In Figure 3, we analyze the relation between fields and techniques. We projected the semantic barycenters of the ML techniques onto the disciplinary map. Techniques are color-coded by their epistemic goal: inference (yellow), prediction (orange), or other (white). Machine learning, neural networks, deep learning and image processing are next to computer sciences. Robot and genetic algorithms are closer to engineering, while AI is between computer sciences and social sciences. Logistic regression is below medicine, and linear regression is between physical sciences and life sciences clusters, while principal component analysis is near the life sciences cluster. The proximity of logistic regression to medicine and neuroscience illustrates that these fields traditionally adopted the data modeling culture (Breiman, 2001) oriented to provide transparent explanations. Techniques such as the linear and logistic models were historically a requirement for legitimacy in these fields where the object of inquiry demands transparency. The results also suggest that inference techniques have been incorporated across fields and are already part of the disciplines. The clustering of deep learning and neural networks near computer sciences and engineering reflects two things: that these are the fields where the techniques are developed and also fields where the scientific authority is derived from prediction performance.

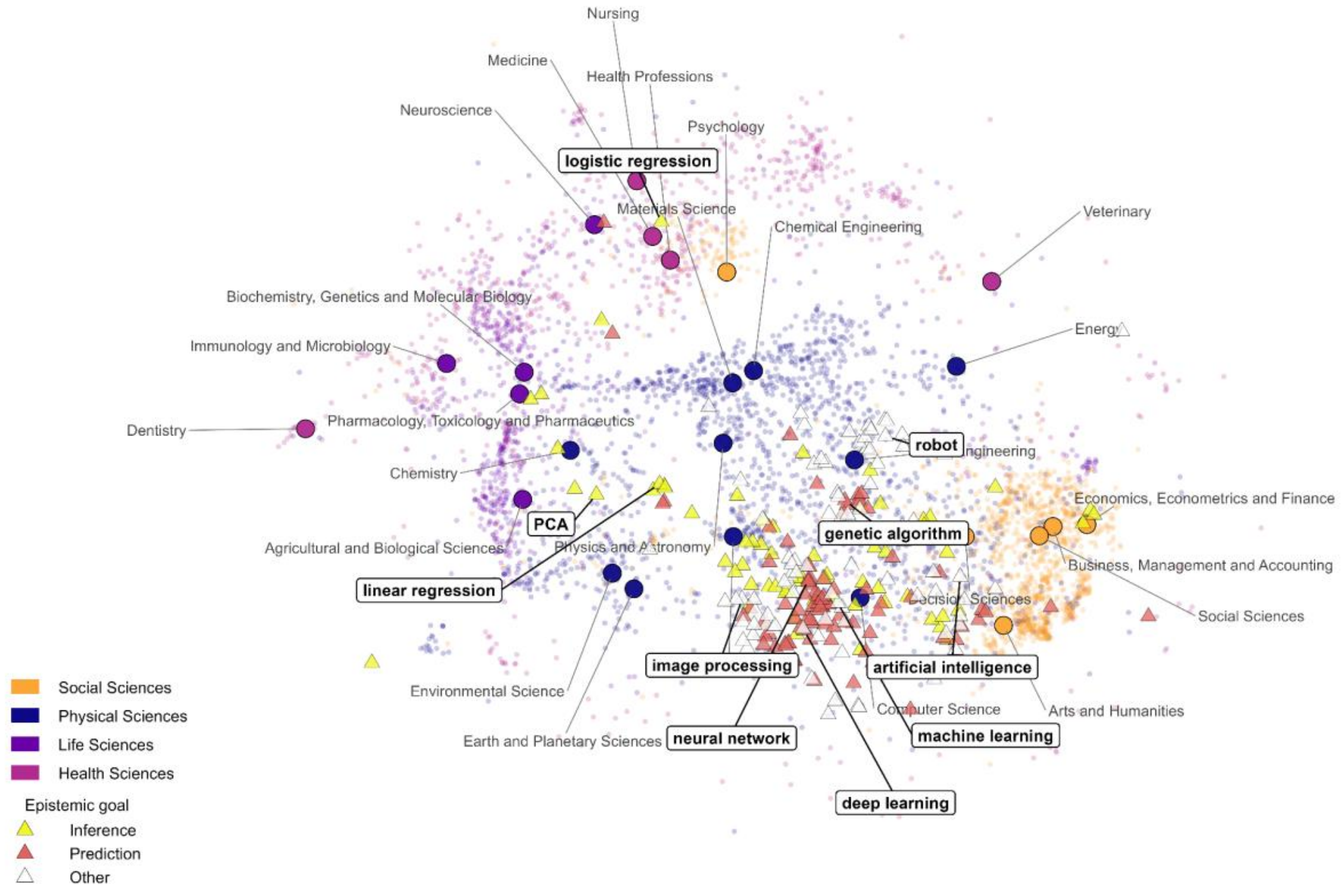

**Figure 3.** ML semantic space: overlay of machine learning techniques onto scientific fields (1990–2025). This map visualizes the distribution of all 255 machine learning techniques. Each triangle represents the semantic barycenter of a technique, color-coded by its epistemic goal (inference, prediction, or other). The top 10 most frequent techniques are labeled. All scientific fields are projected as large points and are fully labeled.

## ML techniques across fields of knowledge

In Figure 4, we compared each technique's share of a field's publications before and after 2015. The clearest pattern belongs to deep learning and convolutional neural networks that exhibit exponential growth across fields, alongside field-specific spikes such as object detection in dentistry. General machine learning shows a consistent adoption that makes it ubiquitous from the chemical sciences to the social sciences. Linear and logistic regression show a bifurcated trajectory. In most disciplines, such as engineering, medicine, and the agricultural sciences, these techniques show a marked contraction in relative prominence, whereas they maintain relative stability or marginal growth within the social sciences, business, economics, and the humanities. Fuzzy logic, genetic algorithms, image processing, and data mining have experienced a shrinkage in relative importance. Overall, Figure 4 shows a reconfiguration of the methodological landscape in favor of deep learning.

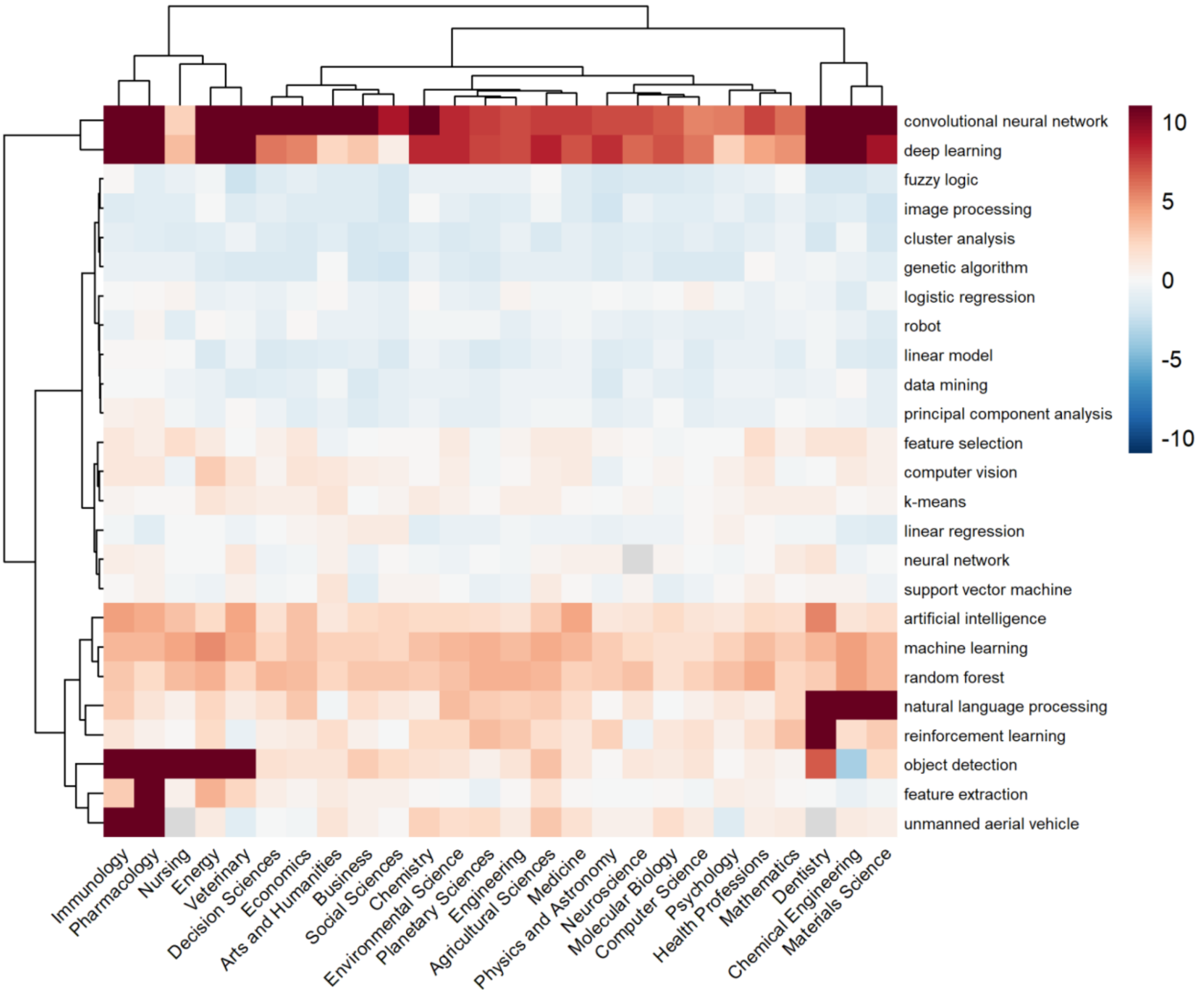

**Figure 4**. Methodological displacement of ML techniques (1990–2025). This heatmap illustrates the shift in techniques' prominence across fields by comparing period 1990–2015 to period 2016–2025. The y-axis lists the top 25 techniques most frequently used in the period 2016–2025. Colors reflect how each technique's share within a field grew or shrank compared to the earlier period. Scale shows red for techniques gaining disciplinary share and blue for those losing ground. Techniques that had no recorded presence in a given field during Period 1990–2015 are assigned the maximum value. Hierarchical clustering highlights shared trajectories.

In the supplementary file, Figure S1 displays the technique–field heatmaps for 1990–2015 and 2016–2025, showing the distribution of technique weights across fields. By visualizing the top 25 techniques for each period in separate heatmaps, we highlight the structural composition in the two time windows.

## Longitudinal trends and the deep learning shift

In Figure 5, we examine which ML techniques are most adopted in the corpus and which have emerged or declined over time. Panel A, with the temporal evolution of the top 10 techniques, illustrates that machine learning is the primary descriptor. Logistic regression and linear regression have undergone a decline in their relative share of the total ML publications since 2016. After 2015 deep learning has grown exponentially driven by a sequence of technical advances that lowered the barriers to training deeper networks. This evolution began in earnest within computer vision when Krizhevsky et al. (2012) demonstrated the potential of convolutional neural networks by winning ImageNet achieving a 15.3% top-5 error rate that shattered the runner-up's 26.2%. This milestone catalyzed natural language processing (NLP), where Mikolov et al. (2013) advanced in early neural language models. The mid-2010s saw a rapid diversification of architectures that pushed the boundaries of network depth and stability. Vaswani et al. (2017) proposed the transformer in 2017, an architecture based solely on attention mechanisms that eliminated the need for convolution. It allowed subsequent models, such as BERT (bidirectional encoder representations from transformers) (Devlin et al., 2018), which produces deep, bidirectional language representations and has set new state-of-the-art records in NLP tasks.

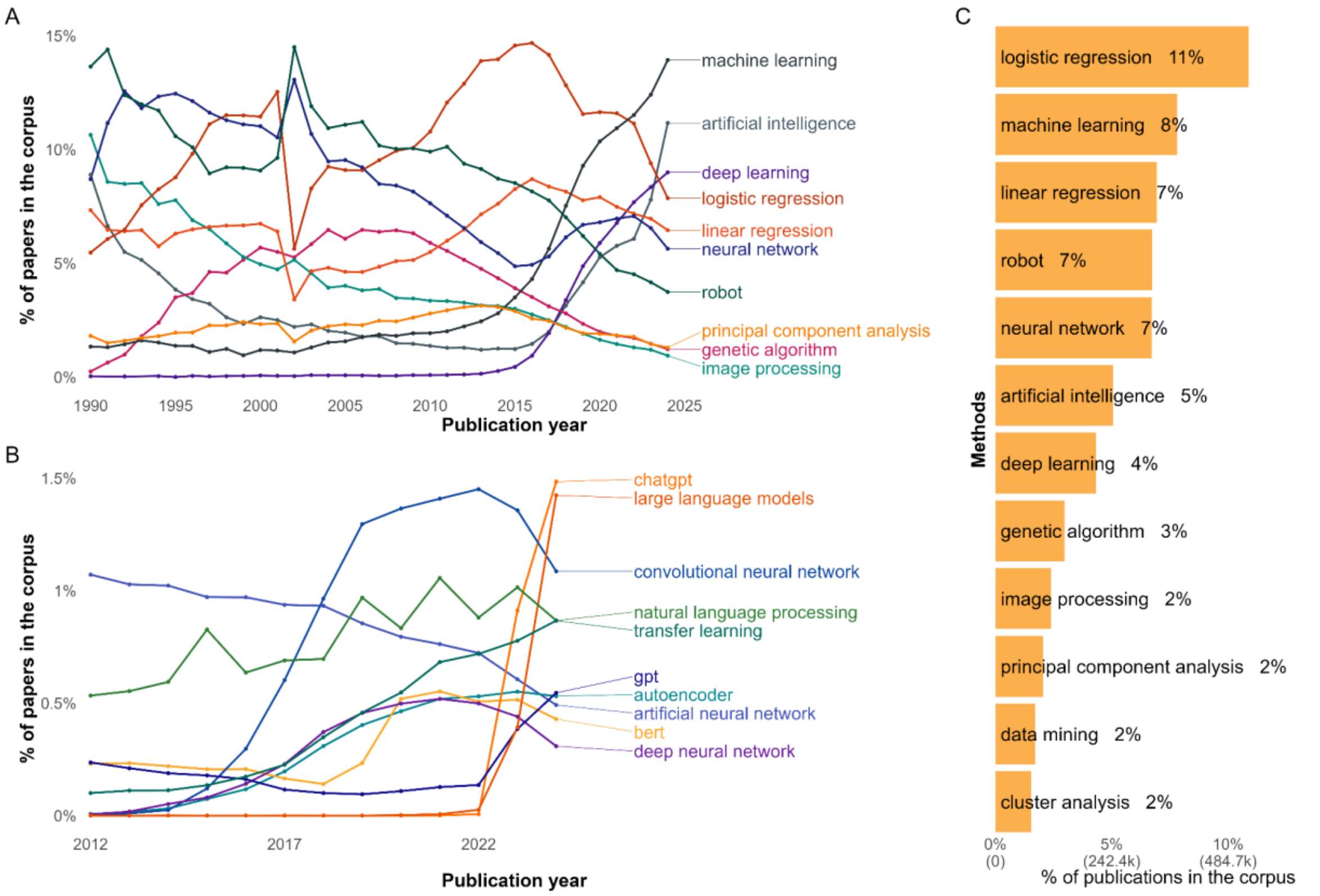

**Figure 5.** Evolution of ML and DL techniques. (A) Temporal trends for the top 10 machine learning techniques (1990–2025). (B) Temporal trends for the top 10 deep learning architectures (2012–2024). (C) Proportion of publications in the corpus for the top 12 techniques (absolute counts are indicated in parentheses). Percentages in panels A and B are computed using document-level fractional counting, where each publication distributes unit weight across its associated keywords. Yearly values are obtained by aggregating these fractional weights and normalizing by the total corpus size per year.

Panel B is focused on deep learning and shows that since 2022 the field has witnessed a second major leap. While the first wave of growth (2015-2021) was characterized by the dominance of convolutional neural networks, the temporal data shows they reached a plateau and subsequent decline after 2021. This saturation effect reflects a change made possible by the transformer architecture mentioned above (Vaswani et al., 2017), which made CNNs redundant for many tasks. Panel B illustrates this transition along a continuum of investigator control: at one end, architectures such as CNN, autoencoder, artificial neural network, and deep neural network are trained by the researcher on their own data and can be reported in a methods section. In the middle, architectures such as BERT, NLP tools, and transfer learning depend on pre-trained representations whose training process is external to the researcher and whose application is controllable and reportable. Large language models, GPT and ChatGPT show the steepest growth trajectory post-2022, architectures where the investigator controls only the prompt, while training data, parameter weights, and model updates remain outside their reach. This shift highlights a transition toward architectures where the researcher's methodological control is decreased with direct consequences for what can be reported, replicated, and independently verified.

Panel C illustrates the relative frequency of specific ML techniques across the entire study period (1990–2025), showing the percentage of papers in the corpus where each technique appears. Logistic regression (11%) is the most frequent technique, followed by machine learning (8%), linear regression (7%), robot (7%), neural network (7%), artificial intelligence (5%) and deep learning (4%). While these top-tier techniques account for the 50% of the corpus, the remaining half is characterized by a heavy-tailed distribution.

Thus, Figure 5 documents a methodological transformation with two distinct phases. The first (2015-2021) displaced interpretable techniques in favor of deep learning architectures that traded transparency for predictive power and remained within the investigator's methodological control. The second (post 2022) introduces a qualitatively different condition: the fastest-growing architectures are no longer tools the researcher trains and reports, shifting the locus of methodological control from the scientific community to external actors. Together, these two phases describe a shift in what techniques science uses, as well as a progressive reorganization of who controls the infrastructure through which scientific claims are produced.

Figure 6 illustrates the evolution of methodological shares across domains, highlighting a direct tension with Breiman's (2001) two cultures. The decline of linear and logistic regression is particularly pronounced in health sciences, life sciences, and social sciences. The shift toward an algorithmic culture prioritizes predictive performance, and as Panel B in Figure 5 shows, increasingly does so through architectures whose training process lies outside the investigator's control.

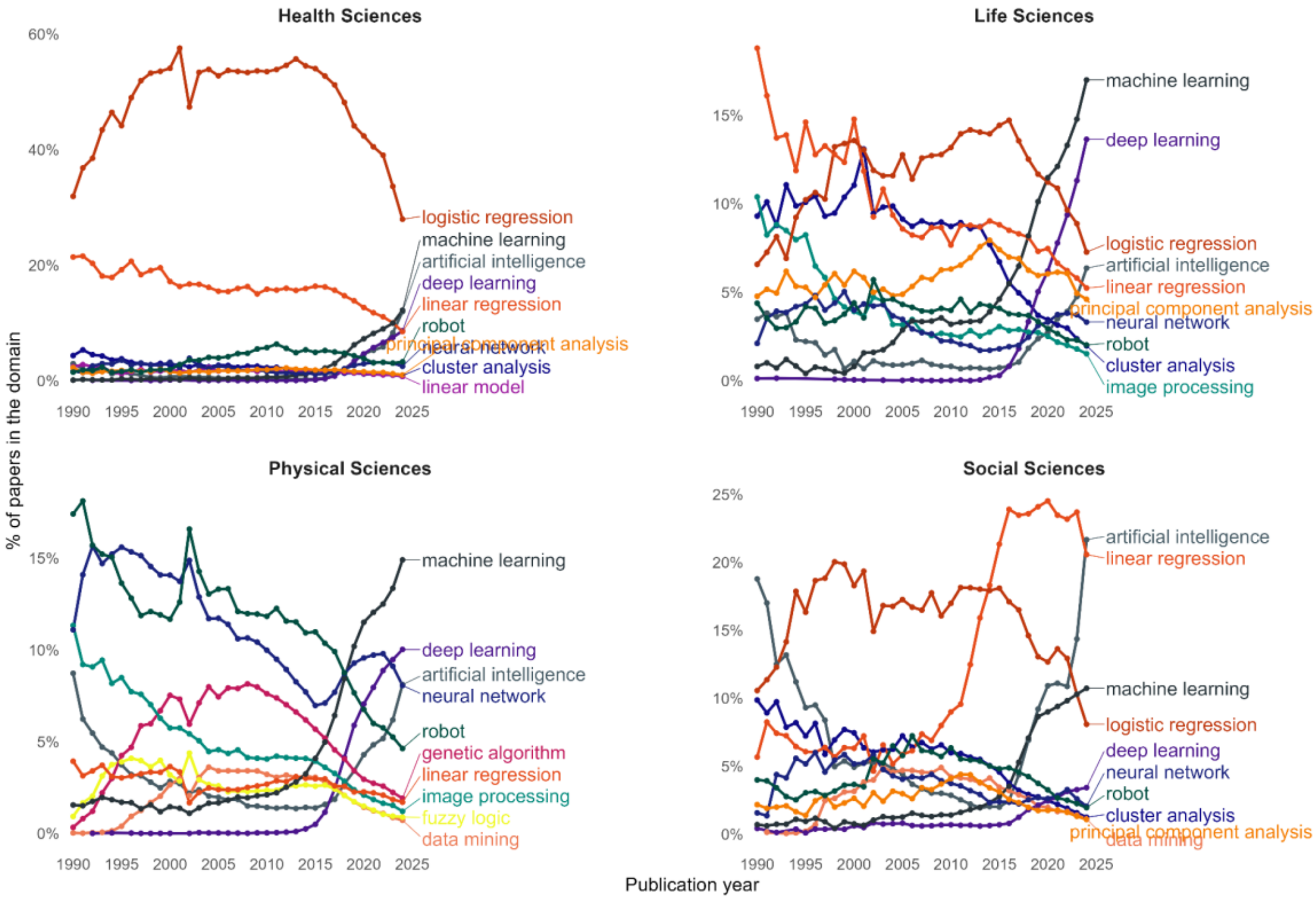

**Figure 6.** Temporal evolution of the top 10 ML techniques across scientific domains (1990–2024). The vertical axis represents the yearly share of each technique within a domain's total output. Percentages are computed using fractional counting to account for multi-label co-occurrence and to reflect relative keyword prominence within articles.

## From inference to prediction

Figure 7 explores whether the decreasing tendency of logistic and linear regression is merely a relative shift within the ML landscape or a broader systemic displacement. By analyzing the relative growth rate (RGR), we observe that since approximately 2015, techniques associated with the algorithmic culture, specifically deep learning and AI, have consistently maintained an RGR > 1. Thus, they are expanding significantly faster than the baseline scientific literature in all domains. In contrast, linear and logistic regressions have largely converged toward or fallen below the growth parity threshold (RGR = 1). In health and social sciences, these once-dominant inferential tools are now growing slower than the disciplinary average showing a shift within the fields that historically required interpretability as a condition for valid knowledge production.

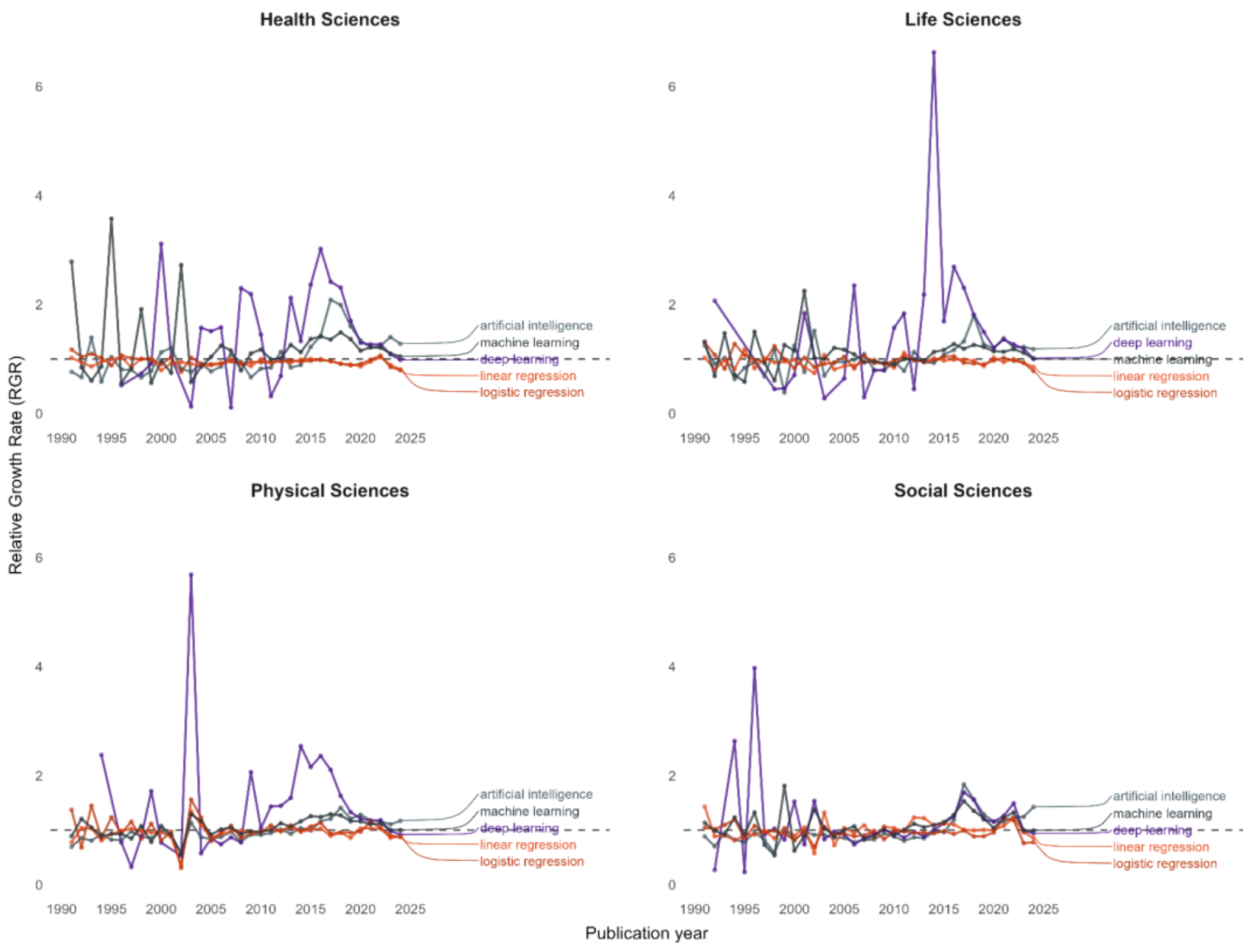

**Figure 7.** Relative growth rate of and ML techniques across scientific domains (1990–2024). The RGR compares the year-over-year growth of a keyword's share within our ML corpus against the baseline year-over-year growth of the entire corresponding scientific domain in the OpenAlex database. The dashed horizontal line at RGR = 1 denotes growth parity. When a line is above this threshold (RGR > 1), it indicates that the keyword's prevalence is growing faster than the overall baseline literature of that disciplinary field. When a curve falls below 1, it indicates the technique is growing slower than the broader domain.

## Discussion

Four research questions motivated this study, related to the evolution, the disciplinary and semantic organization, and the epistemic implications of machine learning across scientific fields. Results show an exponential expansion of ML publications structured around a core-periphery configuration in which physical sciences is the core of the corpus, while health sciences is the primary area of application, followed by social sciences and life sciences. We observe a historically disciplinary differentiation: deep learning architectures are centered in computer sciences, the field that both develops and validates them, whereas logistic regression remains strongly anchored in medicine, reflecting distinct validation regimes across fields. The temporal evolution indicates a decline in techniques oriented to inference, which grow more slowly than their respective disciplinary baselines in OpenAlex, alongside a marked expansion of deep learning architectures across domains.

The epistemological assumptions of a discipline are most directly observable through the techniques it uses, as techniques incorporate the validation criteria that define what legitimate knowledge looks like in each field (Marradi et al., 2007). Techniques reflect disciplinary questions and also constitute the horizon of what can be interrogated (Campolo & Schwerzmann, 2023). When a discipline adopts a new family of techniques, it changes the structure of what can be known within that discipline.

Inference techniques decline in health, life, and social sciences. Figures 6 and 7 show that deep learning and AI have grown faster than the overall scientific output of all four domains since 2015, while logistic and linear regression have stagnated or fallen below baseline growth. Breiman's (2001) algorithmic culture seems to have migrated into disciplines that have historically organized their validation standards around interpretability rather than predictive performance. Thus, legitimacy appears increasingly grounded in predictive performance rather than in the capacity to produce interpretable explanations. This seems to be a change of the epistemic standards beyond production, validation, and legitimation of knowledge (Iliadis & Russo, 2016; Kitchin, 2014).

The diffusion of ML has two opacity layers, one technical opacity referring to the mismatch between human comprehension and mathematical optimization of DL and a second layer due to the commercial secrecy of the models. In the first wave of deep learning diffusion (2015-2021) the models could be trained, reported, and in principle replicated. The rapid uptake of LLM architectures after 2022 visible in Figure 5, Panel B, suggest the emergence of a second layer of opacity. The second wave is organized around a small number of models (Hodel & West, 2025) delivered through a small number of companies, propagating horizontally (Van Der Vlist et al., 2024). Weights, training data, and architectural decisions are mostly defined by actors operating under commercial rather than scientific logics, with no obligation to disclose what would allow independent audit or replication, which leads to what Cruz-Aguilar (2026) calls epistemic enclosure. The historically negotiated criteria for what counts as legitimate knowledge in each discipline now encounters an infrastructure it cannot inspect, and in many cases, cannot even fully describe in a methods section (Bartlett et al., 2018; Binz et al., 2025). When interpretable techniques are displaced in fields where interpretability was a requirement of legitimate knowledge, a discipline can maintain the form of its validation culture while the substance has changed.

## Conclusions

In this study, we analyzed the evolution of machine learning across 4.9 million scientific publications between 1990 and 2025, throughout all scientific domains. By analyzing 255 ML techniques, we observe a general transition from inference to prediction that has permeated fields that have historically defined their epistemic identity through interpretability.

We see that predictive performance becomes a dominant standard, mediated through opaque and, in many cases, proprietary infrastructures (Gillespie, 2024). The rise of machine learning expands the analytical capacity of science, and reorganizes the conditions under which knowledge claims can be produced and evaluated.

This transformation raises concrete challenges for information science. How should peer review evaluate knowledge claims whose internal logic cannot be independently inspected? And how can research communities maintain accountability in a landscape increasingly organized around opaque, proprietary architectures? These are central questions about the organization, validation, and circulation of knowledge in contemporary science.

## Acknowledgments

This project was funded by the Social Science and Humanities Research Council of Canada Pan-Canadian Knowledge Access Initiative Grant (Grant 1007-2023-0001), and the Fonds de recherche du Québec - Société et Culture through the Programme d’appui aux Chaires UNESCO (Grant 338828).